\def\arxiv{}

\documentclass[runningheads]{llncs}

\usepackage{cite}
\usepackage{url}
\usepackage{amsmath}
\usepackage{calc}
\usepackage{algorithm}
\usepackage{algpseudocode}
\usepackage{booktabs}
\usepackage{enumitem}
\usepackage{verbatim}
\usepackage{ulem}
\usepackage{soul,xcolor}
\setstcolor{red}

\algrenewcommand\algorithmicrequire{\textbf{Input:}}
\algrenewcommand\algorithmicensure{\textbf{Output:}}
\usepackage{float}
\makeatletter
\newenvironment{credits}{%
\begingroup\small%
\renewcommand\subsubsection{\@startsection{subsubsection}{3}{\z@}%
        {-12\p@ \@plus -4\p@ \@minus -4\p@}%
        {-0.5em \@plus -0.22em \@minus -0.1em}%
        {\normalfont\small\bfseries\boldmath}}
\renewcommand\paragraph{\@startsection{paragraph}{4}{\z@}%
        {-8\p@ \@plus -4\p@ \@minus -4\p@}%
        {-0.5em \@plus -0.22em \@minus -0.1em}%
        {\normalfont\small\itshape}}%
}{\endgroup}
\makeatother

\usepackage{graphicx}
\usepackage[%
breaklinks=true,
colorlinks=true,
citecolor=blue,
linkcolor=blue,
urlcolor=blue,
]{hyperref}

\usepackage{siunitx}
\usepackage{acro}

\usepackage{siunitx}

\usepackage{xspace}

\newcommand{\haldev}{\textit{ds\_dev\_ill\_hal}\xspace}
\newcommand{\led}{\textit{ds\_dev\_ill\_led}\xspace}
\newcommand{\haltest}{\textit{ds\_test\_ill\_hal}\xspace}
\newcommand{\pigdev}{\textit{ds\_dev\_pig}\xspace}
\newcommand{\pigtest}{\textit{ds\_test\_pig}\xspace}
\newcommand{\rat}{\textit{ds\_test\_rat}\xspace}
\newcommand{\cc}{\textit{ds\_test\_cc}\xspace}
\ifx\arxiv\undefined

\else

\fi

\def\discintname{Disclosure of Interests.}
\def\ackname{Acknowledgments.}

\begin{document}
\title{Deep intra-operative illumination calibration of hyperspectral cameras}
\titlerunning{Deep intra-operative illumination calibration of hyperspectral cameras}

\author{
Alexander Baumann \inst{1,2,7}\and
Leonardo Ayala \inst{2,7}\and
Alexander Studier-Fischer\inst{6,8,9,10} \and
Jan Sellner \inst{2,3,4,5} \and
Berkin Özdemir\inst{6,9,10} \and
Karl-Friedrich Kowalewski \inst{8,9,10} \and
Slobodan Ilic \inst{1} \and
Silvia Seidlitz\inst{2,3,4,5,}\thanks{Silvia Seidlitz and Lena Maier-Hein contributed equally to this work.}\and
Lena Maier-Hein\inst{2,3,4,5,7,\star} 
}

\authorrunning{Baumann et al.}
\institute{
%1
Siemens AG, Munich, Germany \and
%2
Division of Intelligent Medical Systems, German Cancer Research Center (DKFZ), Heidelberg, Germany \and
%3
Helmholtz Information and Data Science School for Health, Karlsruhe/Heidelberg, Germany \and
%4
Faculty of Mathematics and Computer Science, Heidelberg University, Heidelberg, Germany \and
%5
National Center for Tumor Diseases (NCT), NCT Heidelberg, a partnership between DKFZ and university medical center Heidelberg \and
%6
Department of General, Visceral, and Transplantation Surgery, Heidelberg University Hospital, Heidelberg, Germany \and
%7
Medical Faculty, Heidelberg University, Heidelberg, Germany \and
%8
Department of Urology and Urosurgery, University Medical Center Mannheim, Medical Faculty of the University of Heidelberg, Mannheim, Germany \and
%9
German Cancer Research Center (DKFZ) Heidelberg, Division of Intelligent Systems and Robotics in Urology (ISRU), Heidelberg, Germany \and
%10
DKFZ Hector Cancer Institute at the University Medical Center Mannheim, Mannheim, Germany
}

\maketitle              % typeset the header of the contribution
\begin{abstract}
Hyperspectral imaging (HSI) is emerging as a promising novel imaging modality with various potential surgical applications. Currently available cameras, however, suffer from poor integration into the clinical workflow because they require the lights to be switched off, or the camera to be manually recalibrated as soon as lighting conditions change. Given this critical bottleneck, the contribution of this paper is threefold: (1) We demonstrate that dynamically changing lighting conditions in the operating room dramatically affect the performance of HSI applications, namely physiological parameter estimation, and surgical scene segmentation. (2) We propose the first learning-based approach to automatically recalibrating hyperspectral images during surgery and show that it is sufficiently accurate to replace the tedious process of white reference-based recalibration. 
%In capturing the full diversity of lighting conditions potentially encountered in clinical practice, our data-centric approach was designed to be highly generalizable. This was specifically made possible through the disentanglement of the space of possible illuminations and the space of possible tissue configurations as well as the use of simulations.
(3) Based on a total of 742 HSI cubes from a phantom, porcine models, and rats we show that our recalibration method not only outperforms previously proposed methods, but also generalizes across species, lighting conditions, and image processing tasks. Due to its simple workflow integration as well as high accuracy, speed, and generalization capabilities, our method could evolve as a central component in clinical surgical HSI. 
\end{abstract}

\section{Introduction}
\begin{figure}[b!]
    \centering
    \includegraphics[width = \linewidth]{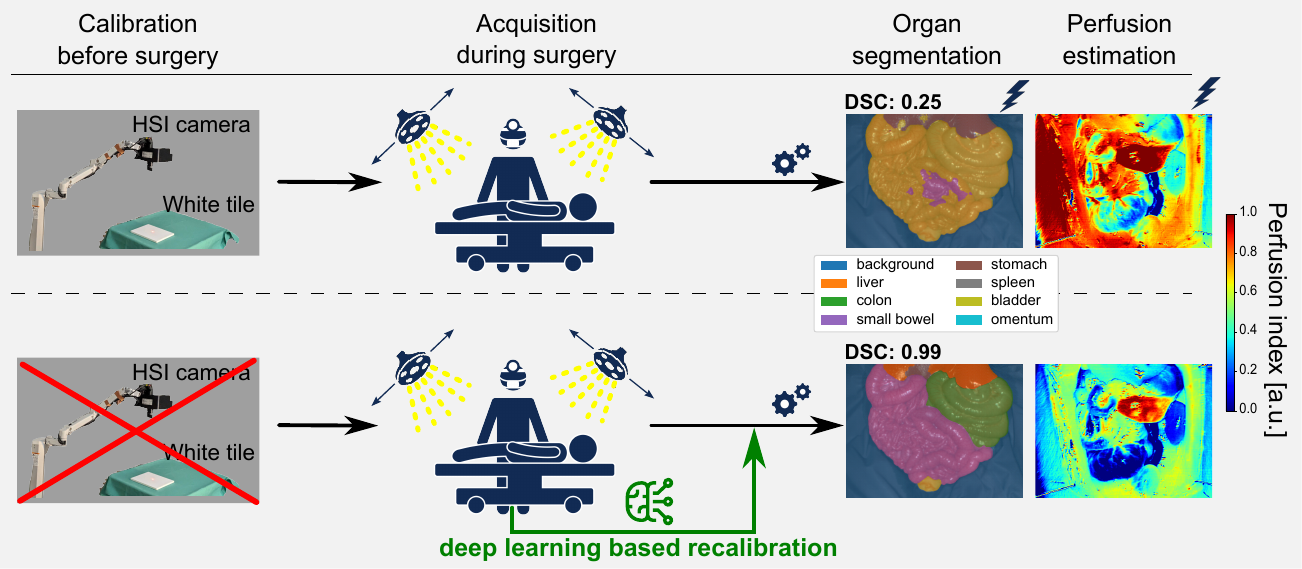}
    \caption{Motivation: Current hyperspectral cameras, which require known lighting conditions, fail in real-world scenarios with dynamically changing lighting conditions.}
    \label{fig:figure_1}
\end{figure}

Hyperspectral imaging (HSI) emerges as a promising medical imaging modality that offers distinct advantages over conventional RGB imaging. In particular, HSI captures spectral information across numerous contiguous bands, thereby enriching the representation of the underlying sample. Recent works have demonstrated the resulting enhancement in tissue classification
\cite{seidlitz2022robust,sellner2023semantic, hsi_tongue_seg, halicek2019vivo, fei},
and the capability of estimating physiological tissue parameters \cite{clancy2020surgical, ayala2019live, ayala2023spectral, kulcke2018compact, holmer2016oxygenation, wirkert2017physiological, shapey2019intraoperative}. 
However, in open surgery, spectral data is affected by changes in illumination and must be correctly calibrated whenever lighting conditions vary \cite{ebner2021intraoperative}. 
The gold standard to achieve this is to switch off all light sources before acquisition \cite{studier2023heiporspectral, kulcke2018compact}. 
As this protocol severely disrupts the clinical workflow, it is presumed not to be consistently applied – leading to unreliable data acquisition and severe failures in downstream tasks, as illustrated in Fig. \ref{fig:figure_1}. This may be one reason why spectral imaging has not yet found widespread use in clinical practice.
%Calibration in surgical spectral imaging is subject to a number of considerable, domain-specific challenges, in particular regarding the dynamic nature of lighting conditions and the high stray light intensity. 
Conventional HSI calibration is conducted with physical white reference measurements, capturing the surrounding illumination. However, they pose challenges in terms of time consumption and sterility, rendering them impractical in the operating room (OR) context. The proposed use of white OR rulers as a sterile alternative \cite{bahl2023synthetic} still presents considerable challenges in the shape of additional workload and their small size.
A number of automatic calibration algorithms originally devised for RGB imaging, such as Gray-world \cite{grayworld} and Max-RGB \cite{land1977retinex}, recover a global illuminant of the scene based on intensity statistics. An alternative calibration strategy tailored specifically to HSI leverages specular highlights for illuminant estimation \cite{ayala2020light}. However, these methods rely on unrealistic assumptions, such as a homogeneous illumination across the entire surgical scene, and may thus fail in practice. Multi-illuminant color constancy models devised for RGB imaging have shown potential in overcoming this issue, as they predict pixel-wise illuminants for calibration \cite{angulargan, gao2019combining, mutimbu2016multiple, hussain2018color, das2021generative}. 
Notably, convolutional neural networks (CNNs) have shown superior performance compared to non-learning-based methods
\cite{angulargan,das2021generative}.
Spectral imaging has so far seen the development of one deep learning approach for multi-illuminant calibration, factorizing reflectance and illumination through an unrolling network \cite{li2021multispectral}. 
However, this research was conducted on outdoor scenes and does not generalize well to the illumination setup in an OR.

Overall, the methods proposed in the literature either remain untested for surgical HSI and/or are conceptually not suitable for spatially resolved calibration. Given this bottleneck, the mission of our work was to develop a new workflow-optimized calibration approach that enables widespread clinical spectral imaging. Our specific contribution is threefold: (1) We are the first to experimentally demonstrate that previously proposed calibration methods fail in in vivo surgical settings. (2) We present the first learning-based approach to performing spatially resolved light recalibration of surgical hyperspectral images. Specifically, we propose to replace conventional physical white reference measurements with a data-driven prediction of the corresponding white tile measurement. This enables a seamless and sterile recalibration process during surgery. (3) Based on the downstream tasks of semantic segmentation and physiological parameter estimation, we show that our recalibration method not only outperforms previous methods, but also generalizes across species, lighting conditions, and image processing tasks.
\begin{figure}[bt]
    \centering
    \includegraphics[width = \linewidth]{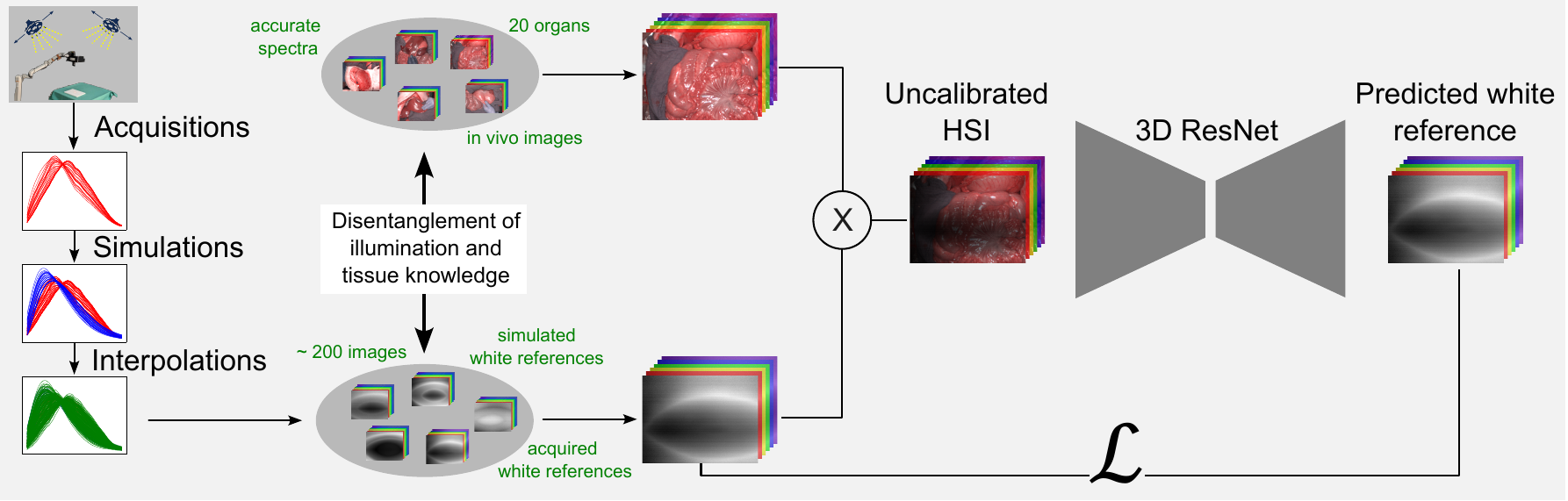}
    \caption{\textbf{The proposed approach replaces tedious manual calibration with a dynamic fully-automatic approach.}
The core of our data-centric method is a 3D-convolutional neural network, trained on in vivo data with artificial light manipulations. At inference time, it takes a raw hyperspectral image as input and generates the corresponding white reference image. The prediction of the white tile image can be used for subsequent calibration of the input image.}
    \label{fig:model}
\end{figure}
\section{Materials and Methods}
The main issue in data-driven calibration is the generalization to unseen settings. To make our method conceptually robust to domain shifts, we propose estimating the white tile measurement that we would obtain for a given scene rather than directly predicting the recalibrated image. Our approach is based on the hypothesis that capturing a representative set of tissue configurations, including illumination conditions, as a training set is infeasible. We therefore disentangle the space of possible illuminations from the space of possible tissue configurations, as illustrated in Fig. \ref{fig:model}. More specifically, we employ a two-dataset training paradigm for the neural network. The first dataset, the illumination dataset, comprises real and simulated white reference images encompassing a wide range of illumination conditions encountered within the OR. The second dataset consists of accurately calibrated HSI cubes of clinically relevant samples. To simulate uncalibrated HSI cubes, each image in the sample dataset is augmented by multiplication with the associated white reference image. Subsequently, the neural network is trained to retrieve the white reference image from the simulated uncalibrated HSI cube. During inference, an uncalibrated HSI cube, possibly acquired with stray light, is fed into the neural network to predict the white reference image needed for illumination calibration.

\subsection{Datasets}
\begin{figure}[tb]
    \centering
    \includegraphics[width = \linewidth]{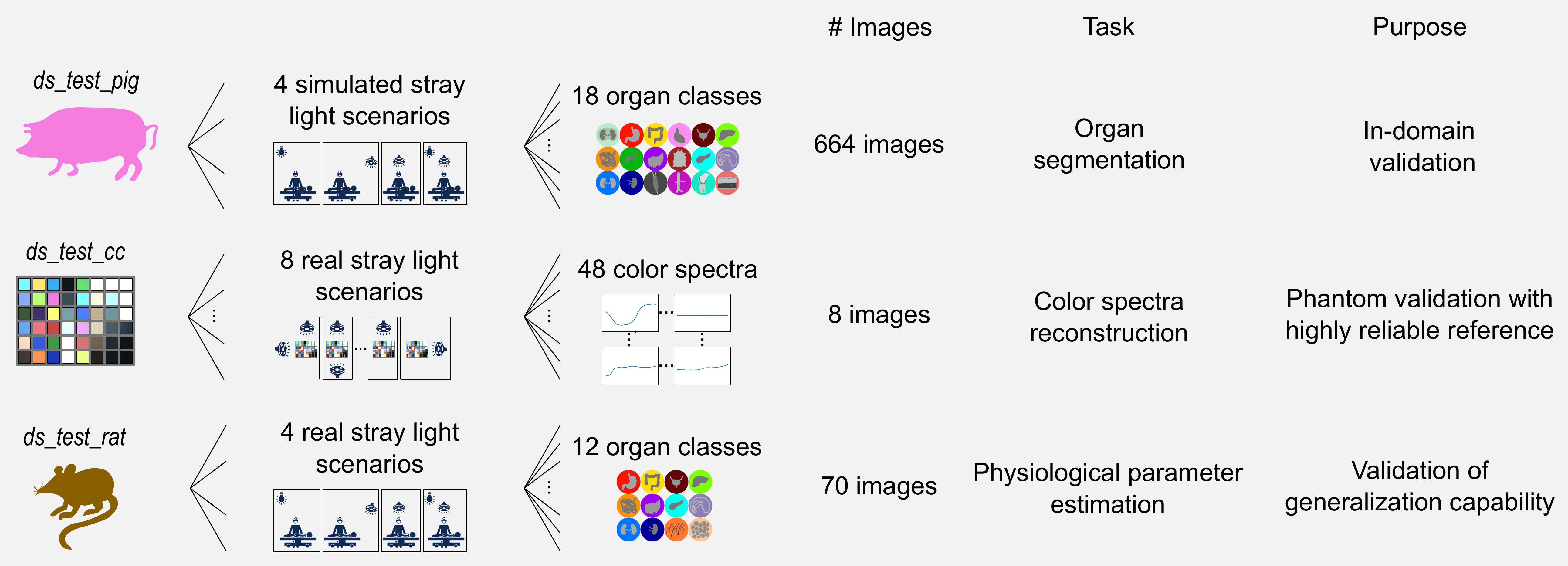}
    \caption{\textbf{Testing concept based on data from three species.}}
    \label{fig:data_fig}
\end{figure}

Training and validation was performed exclusively on porcine data, while testing was performed for unseen stray light scenarios on unseen porcine individuals, a phantom and rats, as summarized in Fig. \ref{fig:data_fig}. 
While the phantom colorchecker board dataset was acquired with the Tivita\textsuperscript{\textregistered} 2.0 Surgery (Diaspective Vision GmbH, Am Salzhaff, Germany) featuring light-emitting diode (LED) illumination, the others were captured with the halogen-based Tivita\textsuperscript{\textregistered} Tissue. As these light sources exhibit different behavior when interfering with the main stray light source, namely LED-based OR lights, we validated on both systems. \\
\indent\textbf{Measured illumination dataset}
The purpose of this dataset was to capture a variety of representative OR lighting conditions for algorithm training. 
To this end, we acquired white reference images in an OR to capture camera light and additional stray light sources such as surgical lights, ceiling lights, or daylight. 
The surgical lights used in our study were manufactured by Dr. Mach and are composed of LEDs (Model: LED 8 MC). 
Diverse stray light scenarios were achieved by varying the angle, distance, and number of surgical lights as well as adjusting blinds or ceiling light, resulting in a wide range of illumination spectra. 
As the two HSI systems used in this study differ in the light sources, we acquired one illumination dataset with each camera:  \led (LED) and \haldev (halogen).\\
\indent\textbf{In vivo porcine development dataset:}
For model development, we curated a subset of the publicly available HSI dataset HeiPorSPECTRAL \cite{studier2023heiporspectral}, \pigtest, consisting of accurately calibrated hyperspectral images of surgical scenes semantically annotated with 18 organ classes.\\
\indent\textbf{Test datasets:}
Comprehensive validation of our methodology was performed based on the three datasets  \pigtest, \cc and \rat summarized in Fig. \ref{fig:data_fig}. 
In-domain testing of calibration quality was performed with data resembling the training data. To simulate stray light in \pigtest, we acquired a white reference test set \haltest with four stray light scenarios (left). 
Colorchecker boards imaged under various lighting conditions were used for the assessment of recalibration performance based on highly reliable reference data (\cc). 
The effect of the recalibration method on surgical image analysis was assessed by means of the downstream tasks organ segmentation and physiological parameter analysis using in-domain and out of domain in vivo hyperspectral imaging data from porcine models (\pigtest) and rats (\rat).

\subsection{Physics-based illumination simulation}
To implement the data-centric recalibration concept, we focused on the model-based generation of plausible white tile data. 
To overcome the resource-intensive white reference acquisitions, we enhanced \led and \haldev by synthesizing white tile images based on real L1-normalized white tile images. \\
\textbf{LED simulations:}
Surgical lights are the main source of stray light in the OR. For our LED-based surgical lights, interference with an LED-based HSI system is approximately constructive and local extrema in the spectrum of the camera light source are preserved. This behavior can be modeled by linear inter- and extrapolations of white reference images. To avoid the generation of duplicates, we first conduct clustering, before combining images of distinct clusters. \\
\textbf{Halogen simulations:}
In contrast to LED spectra, halogen spectra differ in terms of width and the number of local extrema. Consequently, the interference with LED light does not preserve local extrema. To obtain powerful simulations of stray light-affected halogen spectra, we propose to mathematically model the curves. Inspired by Planck’s radiation law, we empirically observed that the following parametric function describes stray light-affected halogen spectra:
\begin{equation*}
   f_{a,b,c,d}(\lambda) = \frac{(a\lambda-b)^3}{\exp (c\lambda-d)-1} 
\end{equation*}
where $f$ denotes the intensity, $\lambda$ the wavelength and $a,b,c,d$ the parameters. Least-square optimization to the mean spectra of \haldev yielded parameter ranges for a,b,c,d so that $f_{a,b,c,d}(\lambda)$ adequately models the illumination conditions captured in \haldev. Increasing the upper bounds of these ranges led to halogen spectra with higher levels of stray light.
To synthesize a hyperspectral image that features realistic spatial variations of intensity from the simulated light spectrum $f_{a,b,c,d}(\lambda)$, we leveraged the acquired images. 
More concretely, a hyperspectral image $I(i,j,\lambda)$ is randomly selected from \textit{ds\_dev\_ill\_hal} and divided by its spatially averaged spectrum $\overline{I}(\lambda)$. 
By multiplication with the simulated spectrum $f_{a,b,c,d}(\lambda)$, we obtain a simulated white reference image $I_s(i,j,\lambda)$ with the simulated spectrum $f_{a,b,c,d}(\lambda)$ as mean spectrum and the spatial variations from $I$.
\begin{equation*}
    I_s(i,j,\lambda) = f_{a,b,c,d}(\lambda)\odot I(i,j,\lambda) \oslash  \overline{I}(\lambda)
\end{equation*}
To further enhance the coverage of illumination conditions, inter- and extrapolation was performed as for the LED simulations. 
Overall, this process yielded a set of about 200 different illumination conditions for each light source.
\subsection{Neural network implementation details}
We feed the HSI cubes into a 3D CNN that employs an autoencoder architecture, utilizing ResNet blocks \cite{resnet} in both the encoder and decoder. 
Two design decisions were particularly important for our method’s success: 
During training, we only optimize the predicted white reference image and not the resulting calibrated sample image. 
Furthermore, we omit skip connections between the encoder and decoder. Both design choices aim to prevent the model from relying heavily on the content of the sample images, instead focusing on learning the illumination information. 
As loss function, we employ the MSE-reconstruction loss between the predicted and original white reference image. Further implementation details are provided in Suppl. Tab. \ref{tab:supp_train_details}. 
\section{Experiments and Results}
We investigated the following research questions (RQs):
  \begin{enumerate}[labelindent=0pt,label={\upshape(\bfseries RQ\arabic*)}, leftmargin=!, labelwidth=\widthof{\ref{RQ3}}]
    \item How do dynamically changing lighting conditions in the OR affect the performance of hyperspectral image analysis algorithms?\label{RQ1}
    \item Are neural networks capable of replacing white tile recalibration of hyperspectral cameras in the OR?\label{RQ2}
    \item  To what extent can neural network-based recalibration mitigate the performance drop of hyperspectral image analysis algorithms under varying lighting conditions?\label{RQ3}
\end{enumerate}  
For all experiments involving our method, we used the same model trained exclusively on porcine data (\pigdev). The generalization to unseen domains (here: colorchecker boards and rats) was investigated on untouched test sets. \\

\indent\textbf{Experiment RQ1:}
We used the traditional approaches Gray-world \cite{grayworld}, Max-RGB \cite{land1977retinex}, and a method based on specular highlights \cite{ayala2020light}, as baselines. Additionally, we integrated a learning-based method by adapting the RGB-calibration framework AngularGAN \cite{angulargan} to hyperspectral imaging. To this end, we trained AngularGAN on \pigdev augmented by the originally acquired white reference images. As downstream tasks, we conducted semantic organ segmentation and physiological parameter estimation on in vivo data. 

For the semantic organ segmentation task, we leveraged the accurately calibrated \pigtest and illumination test set \haltest to obtain four stray light-affected versions of each image in \pigtest. Subsequently, the resulting 664 images were recalibrated by one of the methods, followed by the inference of segmentation masks using a public segmentation model trained on calibrated pig organ images \cite{sellner2023semantic}. As segmentation metrics, the Dice similarity coefficient (DSC) and the normalized surface distance (NSD) were used, as recommended by \cite{maier2024metrics}. 
For physiological parameter estimation, recalibration procedures were applied to \rat, followed by computation of the oxygen saturation, perfusion, hemoglobin, and water index \cite{kulcke2018compact}. To gauge calibration performance, mean absolute errors were calculated between the stray light-affected parameters and reference values derived from images devoid of stray light interference.
For both downstream tasks, the hierarchical structure of the data was respected during aggregation. 

\begin{figure}[tb]
    \centering
    \includegraphics[width = \linewidth]{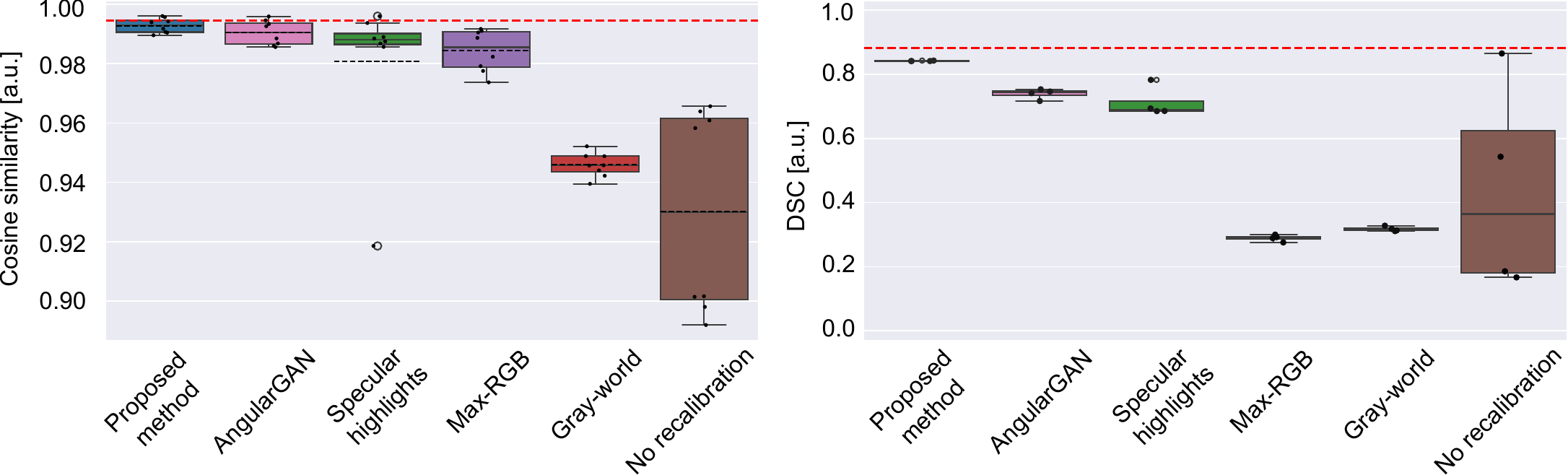}
    \caption{\textbf{State-of-the-art methods fail under dynamically changing light conditions. Our approach addresses this issue.} (Left) Results on colorchecker dataset \cc. The boxplots show the cosine similarity between recalibrated and reference spectra, averaged across colors. Red line: Gold standard of manual white tile calibration. (Right) Results on semantic segmentation dataset \pigtest. Red line: Mean DSC in the absence of stray light. Points: Different stray light scenarios.}
    \label{fig:cc_organseg}
\end{figure}

Fig. \ref{fig:cc_organseg} and Suppl. Fig. \ref{fig:supp_nsd} 
show that existing HSI calibration techniques lack adequate accuracy. Even the best performing methods (Specular highlights and AngularGAN) come with a decrease of the DSC of more than \SI{15}{\percent}.  

\begin{figure}[tb]
    \centering
    \includegraphics[width = \linewidth]{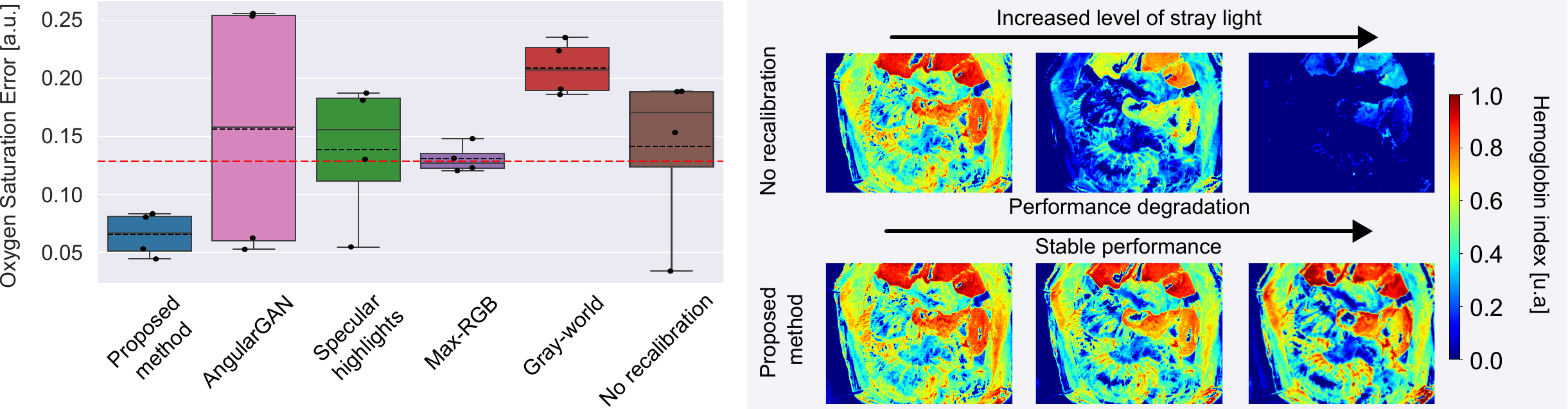}
    \caption{\textbf{In contrast to related methods, our approach generalizes across species.} 
(Left) Organ-specific absolute oxygen saturation errors between calibrated rat images without stray light and corresponding stray light images that are recalibrated by one of the methods. Red line: Mean performance of the gold standard (manual white tile calibration). (Right) Our method yields precise hemoglobin index estimates under dynamically changing lighting conditions.}
    \label{fig:rat_results}
\end{figure}

\begin{figure}
    \centering
    \includegraphics[width=0.8\linewidth]{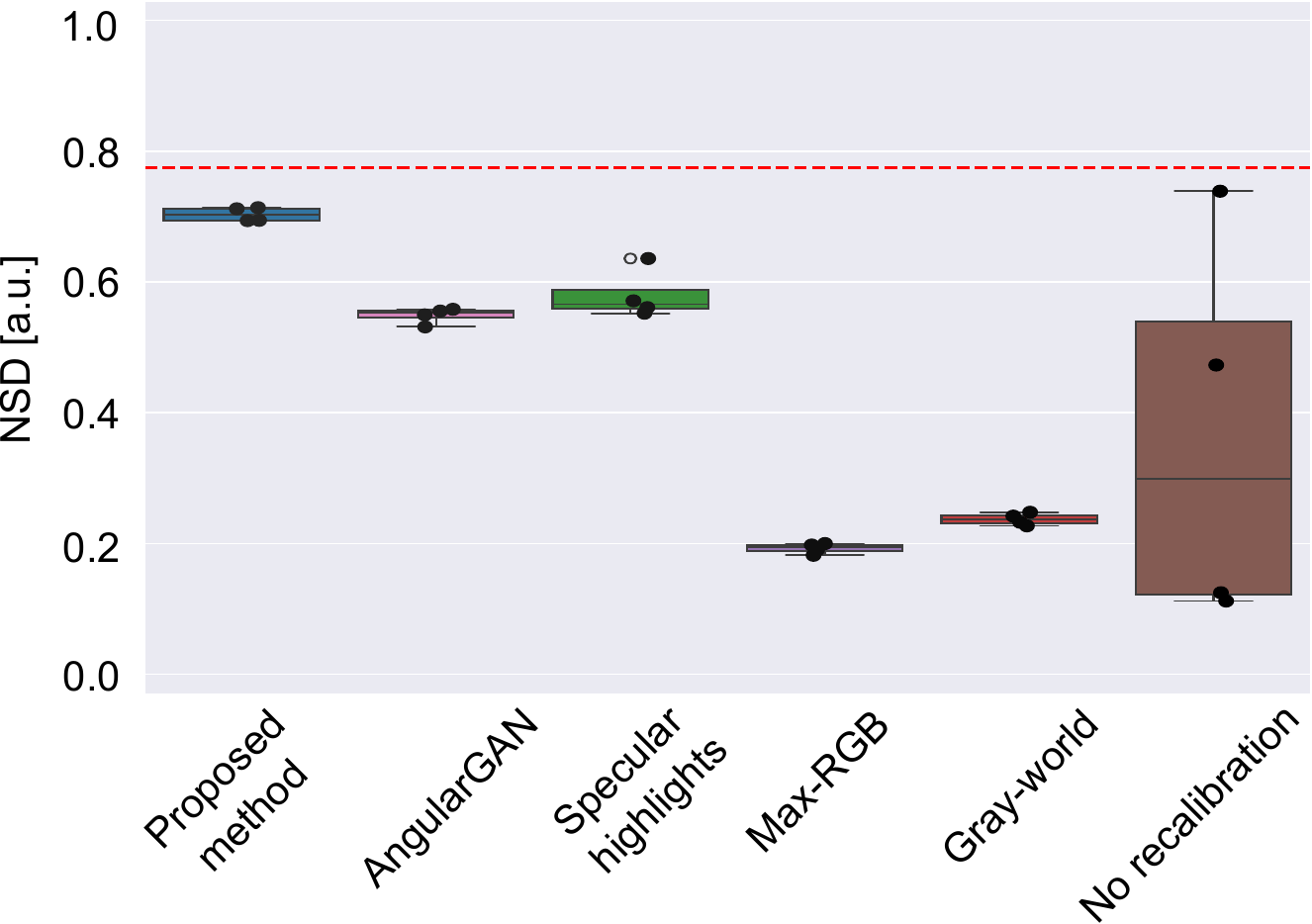}
    \caption{Results on the semantic segmentation dataset. Red line: Mean normalized surface dice (NSD) in the absence of straylight. Points: Different stray light scenarios.}
    \label{fig:supp_nsd}
\end{figure}

\begin{figure}
    \centering
    \includegraphics[width=1\linewidth]{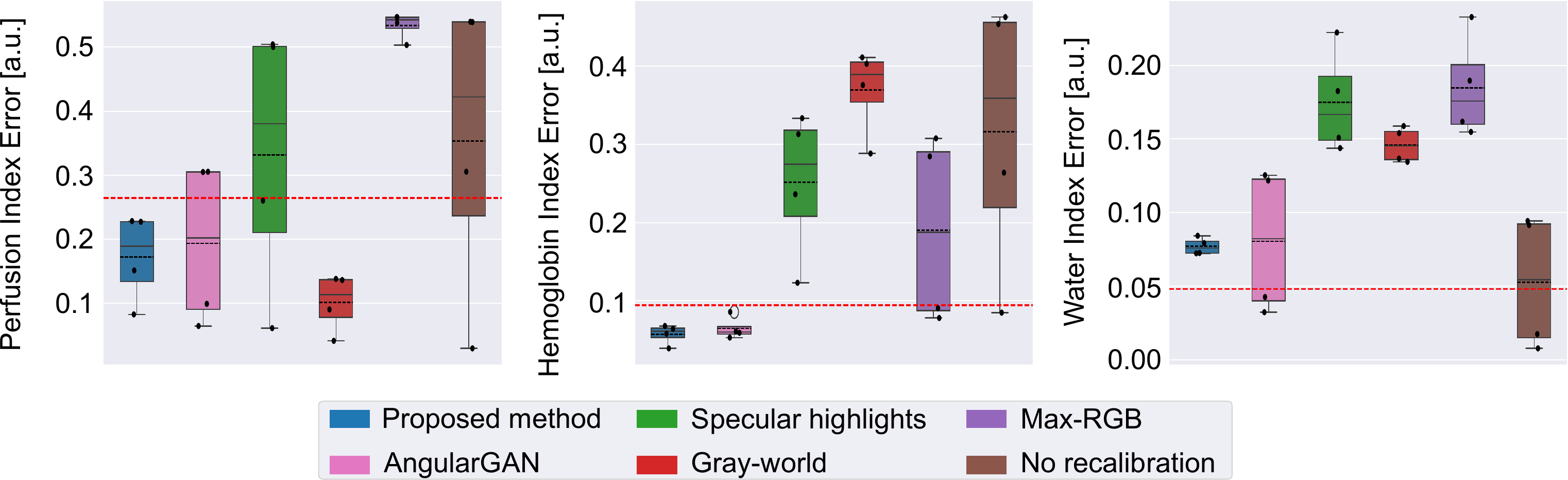}
    \caption{Organ-specific absolute perfusion (left), hemoglobin (middle) and water (right) index errors between calibrated rat images without stray light and corresponding stray light images that are recalibrated by one of the methods. Red line: Mean performance of the gold standard (manual white tile calibration).}
    \label{fig:supp_params}
\end{figure}

Similarly (Fig. \ref{fig:rat_results}), the tissue parameter maps change substantially under varying lighting conditions, even when manual white tile recalibration was performed (cf. Suppl. Fig. \ref{fig:supp_params}). \\%\ref{fig:supp_params}).

\indent\textbf{Experiment RQ2:}
To measure the calibration accuracy in an OOD setting with a highly reliable reference, we applied our model trained on porcine data to recalibrate \cc. As illustrated in Fig. \ref{fig:cc_organseg}, our method demonstrates the highest average cosine similarity. 
They are almost on par with the gold standard of manually acquiring white tile measurements. \\

\begin{figure}
    \centering
    \includegraphics[width=\linewidth]{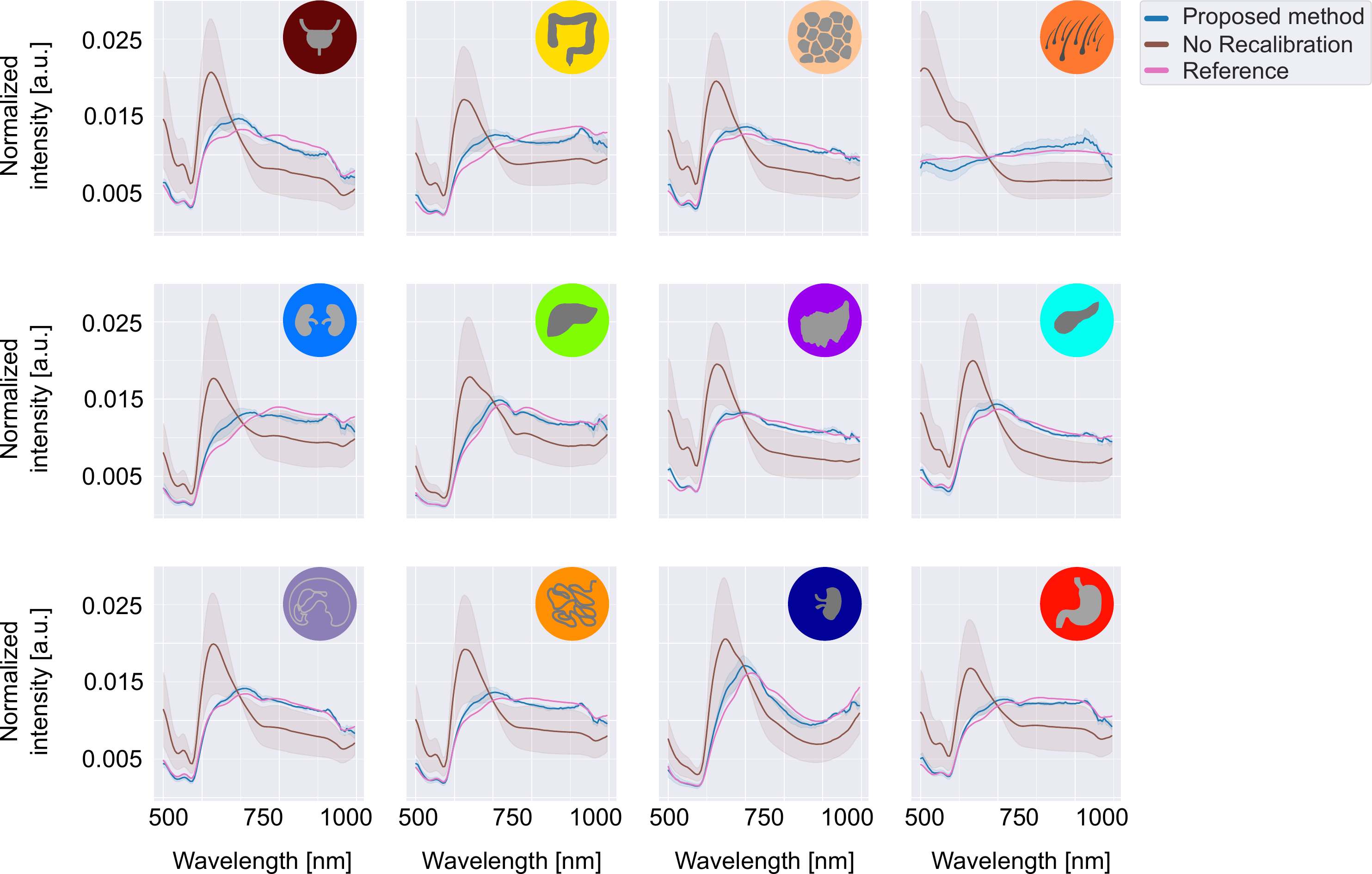}
    \caption{While tissue spectra in the presence of stray light (brown) significantly deviate from reference spectra in the absence of straylight (magenta), our recalibration approach accurately restores the tissue spectra (blue). Solid lines denote the mean L-1 normalized reflectance. Shaded areas depict the standard deviation interval.
}
    \label{fig:supp_organ_spectra}
\end{figure}

\indent\textbf{Experiment RQ3:}
To assess the capability of our method to boost the downstream task performance, we performed Experiment RQ1 on our recalibration approach.
As shown in Figs. \ref{fig:cc_organseg} and \ref{fig:rat_results}, our method outperforms previous methods by a large margin. For semantic segmentation, relative improvements of the DSC from \SI{14}{\percent} to \SI{191}{\percent} were obtained. For oxygen saturation estimation, the error could be reduced by \SI{50}{\percent} to \SI{69}{\percent}. Similar performance gains were obtained for other tissue parameters (cf. Suppl. Fig. \ref{fig:supp_params}). A qualitative assessment of the high fidelity of our recalibrated tissue spectra is available in Suppl. Fig. \ref{fig:supp_organ_spectra}.%\ref{fig:supp_params}). 

\section{Discussion}
We were the first to provide in vivo evidence that dynamically changing lighting conditions in the OR can cause dramatic failures in HSI analysis. This is a finding of high clinical relevance because the manual white tile-based recalibration of cameras during surgery severely disrupts the clinical workflow and may currently hinder widespread clinical adoption of HSI cameras. 
According to our comprehensive study, our method represents the only calibration model that is capable of maintaining high accuracy independently of the downstream task and domain, indicating high applicability for clinical use cases. It also features several major conceptual advantages: White reference measurements are not only impractical as they suffer from sterilization and workflow issues, but are also prone to oversaturation. This explains the suboptimal performance on the rat data (see Fig. 6).
%– as our model is trained on non-oversaturated white reference images, it does not suffer from this problem. 
Specular highlights and Max-RGB exhibit high calibration accuracy on the colorchecker board, as both methods calibrate the images with the white color field by design, but fail to generalize to in vivo scenarios. Note that these methods recover a global illumination estimate, which is not sufficient in the case of spatially heterogeneous illumination encountered in the OR. In fact, we also saw drops in performance when reducing the estimations of white tile calibration (classic and data-driven) to a global estimate. 
Overall, the core strength of our approach is its generalizability.  Notably, it outperforms the competing neural network method AngularGAN even when trained on the exact same data as our method. We attribute this to the inherent domain shift of auto-encoded images. 

A limitation of our work could be seen in the fact that we did not cover all possible illumination settings that can occur in practice. However, as we focused on the most important light sources (surgical lights and ceiling light) and conducted our validation on highly diverse datasets, we are confident that our conclusions will hold in diverse settings.

In conclusion, our work presents the first learning-based light calibration method for hyperspectral imaging. The proposed methodology not only outperforms previously proposed approaches in various settings but can be seamlessly incorporated into hyperspectral imaging systems for ORs. Our work could therefore pave the way for clinical workflow-optimized and robust HSI in surgery. 
\begin{credits}
\subsubsection{\ackname} 
This project was supported by the European Research Council (ERC) under the European Union’s Horizon 2020 research and innovation programme (NEURAL SPICING, 101002198), the National Center for Tumor Diseases (NCT), Heidelberg’s Surgical Oncology Program, the German Cancer Research Center (DKFZ), and the Helmholtz Association under the joint research school HIDSS4Health (Helmholtz Information and Data Science School for Health). We also acknowledge the support through state funds for the Innovation Campus Health + Life Science Alliance Heidelberg Mannheim from the structured postdoc program for Alexander Studier-Fischer: Artificial Intelligence in Health (AIH).
\subsubsection{\discintname}
The authors have no competing interests to declare that are relevant to the content of this article. 
\end{credits}

\newpage
\bibliographystyle{splncs04}
\bibliography{references}

\newpage
\setcounter{section}{0}
\renewcommand{\thesection}{\Alph{section}}
\section{Supplementary material}
\begin{table}
    \centering
    \begin{tabular}{|p{7cm}|p{6cm}|}
        \hline
        \textbf{Optimization algorithm} & Adam Optimizer \\ \hline
        \textbf{Learning rate} & Decay from $10^{-4}$ to $10^{-5}$ \\ \hline
        \textbf{Augmentations during training} & Rotating, flipping \\ \hline
        \textbf{Number of parameters} & 7.7 million \\ \hline
        \textbf{Number of channels in backbone} & 512 \\ \hline
        \textbf{Batch size during training} & 5 \\ \hline
        \textbf{Training time} & 4 hours \\ \hline
        \textbf{Software} & PyTorch 2.0.1 \\ \hline
        \textbf{Hardware} & NVIDIA GeForce RTX 3090 GPU \\ \hline 
    \end{tabular}
    \vspace{5pt}
    \caption{Training details}
    \label{tab:supp_train_details}
\end{table}

\end{document}